\documentclass[11pt,english]{article}
\usepackage[T1]{fontenc}
\usepackage{color}
\usepackage{amsmath}
\usepackage{amssymb}
\usepackage{float}
\usepackage{setspace}
\usepackage{esint}
\usepackage{multirow}
\makeatletter
\usepackage{physics}
\usepackage{amsthm}
\usepackage{graphicx}
\usepackage{amsfonts}
\usepackage{slashed}
\usepackage{afterpage}
\usepackage{cite}
\usepackage{caption}
\usepackage{subcaption}
\usepackage{arydshln}
\usepackage{color}
\usepackage{bm}
\definecolor{darkgreen}{cmyk}{1,0,1,0}

\usepackage{stackengine}

\stackMath

\usepackage[margin=1in]{geometry}

\usepackage{babel}

\makeatother

\usepackage{babel}
\begin{document}
	\title{The standard model, the Pati-Salam model, \\ and "Jordan geometry"}
	\author{Latham Boyle$^a$ and Shane Farnsworth$^b$}
	\maketitle
	 \begin{center}
   $^a$Perimeter Institute for Theoretical Physics, Waterloo, Ontario, Canada \\
   $^b$Max Planck Institute for Gravitational Physics (Albert Einstein Institute), Potsdam, Germany
	\end{center} 
	\begin{abstract}
    We argue that the ordinary commutative and associative algebra of spacetime coordinates (familiar from general relativity) should perhaps be replaced, {\it not} by a noncommutative algebra (as in noncommutative geometry), but rather by a Jordan algebra (leading to a framework which we term "Jordan geometry").  We present the Jordan algebra (and representation) that most nearly describes the standard model of particle physics, and we explain that it actually describes a certain (phenomenologically viable) extension of the standard model: by three right-handed (sterile) neutrinos, a complex scalar field $\varphi$, and a $U(1)_{B-L}$ gauge boson which is Higgsed by $\varphi$.  We then note a natural extension of this construction, which describes the $SU(4)\times SU(2)_{L}\times SU(2)_{R}$ Pati-Salam model.  Finally, we discuss a simple and natural Jordan generalization of the exterior algebra of differential forms.
    \end{abstract}
		\tableofcontents
		
	\section{Introduction}
	\label{Sec_Introduction}
	
In this paper, we will try to take the standard model at face value (without imagining the existence of many additional unobserved fields at high energies), and propose a possible interpretation of what it may be telling us.  In particular, we will assume that the standard model fermions (including three right-handed neutrinos) are the {\it only} fundamental fermions; and we will {\it try} to assume the analogous thing about the standard model bosons, but our interpretation will lead us to introduce two more bosons: a $U(1)_{B-L}$ gauge boson, and a complex scalar field $\varphi$, which is responsible for Higgsing $U(1)_{B-L}$.   In brief, we will motivate the idea that spacetime may, roughly speaking, have "discrete" extra dimensions, described by a certain finite-dimensional Jordan algebra of Hermitian matrices.

To briefly introduce our line of thought, let us begin by considering two long-standing goals in fundamental physics:
\begin{itemize}
\item Goal 1: to unify gravity with the other bosonic (and fermionic?) fundamental fields; and
\item Goal 2: to unify gravity with quantum mechanics.
\end{itemize}
For each of these two goals, we will review a well-known strategy for achieving it, and then suggest a modification of that strategy.

First consider Goal 1.  Here a well known strategy is:
\begin{itemize}
\item	Strategy 1: Ever since Einstein's discovery that gravity corresponds to the geometry of 4-dimensional spacetime, a natural dream has been to unify the gravitational field with the various other fundamental fields by interpreting them all as describing the geometry of some appropriately {\it extended} spacetime.  

The most well-known implementation of this strategy is:
\begin{itemize}
\item Strategy 1A: the Kaluza-Klein approach (as still incorporated in various modern proposals, including string theory), which imagines that spacetime has the form ${\cal M}={\cal M}_{4}\times {\cal M}_{int}$ -- {\it i.e.}\ it is the product of the ordinary 4D spacetime manifold ${\cal M}_{4}$ that we observe, and another "internal" spacetime ${\cal M}_{int}$ (which is {\it also} assumed to be an ordinary smooth manifold) which we do not yet observe.

Unfortunately, there are drawbacks to this strategy.  For one thing, it predicts an infinite collection of 4D fields (in contrast to the finite handful of fields that we actually observe).  To rescue the idea, one is forced into the awkward situation of assuming that the internal space ${\cal M}_{int}$ is unobservably tiny (much smaller than the electroweak scale), so that nearly all of the predicted 4D fields are hidden at unobservably high energies.  Moreover, stabilizing ${\cal M}_{int}$ at this tiny size is no easy feat.  The most well-known stabilization mechanism \cite{Kachru:2003aw} is rather complicated, remains controversial \cite{Obied:2018sgi}, and, if correct, apparently predicts an enormous "landscape" of different vacua.  While this story {\it might} turn out to be right, it seems baroque, and it is natural to hope for a more economical picture.

At its root, the problem with this strategy is tighty linked to the assumption that ${\cal M}_{int}$ is a smooth manifold or, correspondingly, that the algebra ${\cal A}_{int}$ of functions on ${\cal M}_{int}$ (the algebra from which that the coordinate functions are drawn) is $\infty$-dimensional: ${\cal A}_{int}=C_{\infty}({\cal M}_{int})$.  This is ultimately why this picture predicts an infinite number of 4D fields, and why ${\cal M}_{int}$ has so many potentially unstable continuous deformations.
\end{itemize}
This suggests a modified strategy:
\begin{itemize}
\item Strategy 1B: replace the smooth manifold ${\cal M}_{int}$ (at least at an effective level) by some sort of discrete or "finite" space ${\cal M}_{F}$, so that the $\infty$-dimensional algebra ${\cal A}_{int}=C_{\infty}({\cal M}_{int})$ is replaced by a finite-dimensional algebra ${\cal A}_{F}$, and the predicted collection of 4D fields becomes finite.

What sort of discrete/finite space do we want?  One's first thought might be to take ${\cal M}_{F}$ to be a naive discretization of ${\cal M}_{int}$ ({\it e.g.}\ to replace a smooth torus ${\cal M}_{int}$ by a corresponding periodically identified lattice ${\cal M}_{F}$, so that the ordinary $\infty$-dimensional commutative algebra of functions over ${\cal M}_{int}$ is simply replaced by the ordinary finite-dimensional commutative algebra of functions over the lattice ${\cal M}_{F}$).  But this option (called "deconstruction" in the literature \cite{ArkaniHamed:2001ca, Hill:2000mu}) is not what we want, since the algebraic/symmetry structure of the standard model is not properly reflected in the geometry of such an internal space.  What should we do instead?
\end{itemize}
For a further clue, consider Goal 2 (unifying gravity and quantum mechanics).  In this case, a well known strategy is:
\item Strategy 2: A common expectation has been that unifying gravity with quantum mechanics will involve replacing the classical geometry of spacetime (or, relatedly, configuration space) by some type of "quantum geometry."

A well-known implementation of this strategy is:
\begin{itemize}
\item Strategy 2A: "Noncommutative geometry" ("NCG").  In the classical geometry of spacetime (or configuration space), the coordinates functions are drawn from the commutative algebra ${\cal A}$ of functions over the manifold.  NCG is a proposed generalization in which ${\cal A}$ can be noncommutative.

But, once again, something seems wrong with this approach: NCG is the natural quantum generalization of the sympectic geometry of {\it phase space}, not the (Riemannian) geometry of configuration space.  After all, in ordinary quantum mechanics, it is the phase space coordinates which fail to commute ($[q_{i},p_{j}]=i\hbar\delta_{ij}$), while the configuration space coordinates do commute ($[q_{i},q_{j}]=0$).\footnote{There is a sense in which spacetime coordinates may effectively fail to commute in the presence of external (gravitational or electromagnetic) fields, but this is a separate issue, since we are attempting to give a geometric re-interpretation to the standard model, even in flat space, and in the absence of external fields.}  

In quantum mechanics, we deal with operators on Hilbert space.  We can split the space of such operators into two parts, by splitting each operator ${\cal O}$ into its Hermitian and anti-Hermitian parts:
\begin{subequations}
  \begin{eqnarray}
    H&=&\frac{1}{2}({\cal O}+{\cal O}^{\dagger}), \\
    A&=&\frac{1}{2}({\cal O}-{\cal O}^{\dagger}).
  \end{eqnarray}
\end{subequations} 
Note that the (anti-)Hermitian part does not form an algebra under the usual associative product (composition of operators), since the composition of two (anti-)Hermitian operators is not (anti-)Hermitian.  Instead, the anti-Hermitian operators close under the commutator product $[A_{1},A_{2}]$ to form a {\it Lie algebra}, while the Hermitian operators close under the anti-commutator product $\frac{1}{2}\{H_{1},H_{2}\}$ to form a {\it Jordan algebra}.  (Appendix A includes a brief introduction to Jordan and Lie algebras.)

In the usual formulation of quantum mechanics, one works with the full algebra $\tilde{{\cal A}}$ of operators on Hilbert space -- an associative-but-noncommutative algebra, with the product given by operator composition/matrix multiplication.  But, as a consequence, most of the elements in the algebra are unobservable (non-Hermitian) operators.  Why carry around all this excess, unobservable baggage?  Jordan showed that quantum mechanics could be elegantly reformulated, purely in terms of the commutative-but-nonassociative Jordan algebra of observable (Hermitian) operators.\footnote{For a brief introduction to Jordan's formulation of quantum mechanics, see \cite{Townsend:2016xqf}.}  
\end{itemize}
Inspired by this, we are led to:

\begin{itemize}
\item Strategy 2B: "Jordan geometry:"  Instead of taking the algebra ${\cal A}$ of spacetime coordinates to be a noncommutative algebra of operators (with the associative product given by operator composition/matrix multiplication), we will take it to be a Jordan algebra of {\it Hermitian} operators (with the product given by the anti-commutator).  And, in particular, we will take our "discrete" extra dimensions to be described by a finite-dimensional Jordan algebra ${\cal A}_{F}$.

In fact, although the possibility that this geometry has quantum mechanics built in to its foundations is certainly very appealing, we were originally led to Jordan geometry by a completely {\it different} line of argument that had nothing to do with the quantum mechanical considerations mentioned above.  This second line of argument (explained in the next two sections) was based on following a number of hints which came from thinking about previous attempts to describe the standard model of particle physics within the framework of NCG.  Thus, in principle, it may even turn out that our Jordan geometry is not quantum mechanical at all, and is instead a generalized type of {\it classical} geometry, which still needs to be quantized (as explored in the NCG setting {\it e.g.}\ in \cite{Rovelli:1999fq, Hale:2000xc, Besnard:2007hf}).

From a mathematical standpoint, Jordan geometry seems like it may be a very rich and interesting generalization of ordinary (commutative) geometry.  For example, in Appendix A, we explain how it suggests a particularly natural generalization of the exterior algebra of differential forms.  
\end{itemize}
\end{itemize}

The outline of the paper is as follows.  In Section 2, we briefly review the description of the standard model of particle physics in noncommutative geometry (but with an elegant new presentation, due to Barrett, 
building on earlier work by D'Andrea and Dabrowski).
In Section 3, we present our new Jordan formulation of the $SU(3)\times SU(2)\times U(1)$ standard model, and explain why it actually predicts a certain (phenomenologically-viable) extension by $U(1)_{B-L}$.  In Section 4, we show how this, in turn, suggests a natural further extension: a Jordan formulation of the $SU(4)\times SU(2)_{L}\times SU(2)_{R}$ Pati-Salam model.  In Section 5, we discuss a few interesting topics for the future.  
Finally, in Appendix A, we discuss (graded) Lie and Jordan algebras, and their relationship; and we describe a natural Jordan generalization of the exterior algebra of differential forms (as a step toward the formulation of Jordan geometry, and Jordan gauge theory).  Aspects of Jordan geometry and Jordan gauge theory are further developed in a follow-up paper \cite{Farnsworth:2020ozj}.  
	
\section{The standard model from noncommutative geometry}

To help introduce the next section's Jordan formulation of the standard model, it will be helpful to begin in this section by briefly reviewing the NCG formulation of the standard model, which began with the early work \cite{DuboisViolette:1988ir, DuboisViolette:1988ps, DuboisViolette:1989at, DuboisViolette:1988vq, Connes:1990qp}.  Since this is not a paper about NCG, this section is not intended as a proper introduction to the subject, but merely a brief sketch to orient the unfamiliar reader.  Those who wish to know more are pointed to some relevant references along the way.

The NCG formulation of the standard model is presented in the spectral triple framework \cite{Connes:1996gi} (see \cite{vanSuijlekom:2015iaa} for a more detailed introduction).  A "spectral triple" is a proposed generalization of an ordinary Riemannian manifold (or, more precisely, a Riemannian {\it spin} manifold -- {\it i.e.}\ a Riemannian manifold on which spinor fields may be defined).  In ordinary Riemannian geometry, a geometry is specified by giving the pair $\{{\cal M},g\}$: {\it i.e.} the manifold ${\cal M}$ and the metric $g$ on that manifold.  In NCG, one instead gives a "spectral triple" $\{{\cal A},{\cal H},D\}$ consisting of
\begin{itemize}
\item ${\cal A}$: an algebra (thought of as a generalization of the algebra of coordinates on ${\cal M}$);
\item  ${\cal H}$: a Hilbert space (thought of as a generalization of the Hilbert space of spinors on ${\cal M}$) on which ${\cal A}$ is represented as a left-right bimodule; and
\item $D$: a Hermitian operator on ${\cal H}$ (thought of as the generalization of the Dirac operator on ${\cal M}$).
\end{itemize}
There can also be two other ingredients (the chirality operator $\gamma$, which generalizes $\gamma^{5}$, and the anti-linear operator $J$ which maps each spinor to its charge conjugate), as well as a list of axioms and assumptions governing how these various ingredients behave and interact with one another.  

In a series of papers \cite{Boyle:2014wba, Farnsworth:2014vva, Boyle:2016cjt}, we explain how these various ingredients, axioms and assumptions may be formulated in a more succinct and unified way by combining the algebra ${\cal A}$ and Hilbert space ${\cal H}$ into a single super-algebra ${\cal B}={\cal A}\oplus{\cal H}$, with even/bosonic part ${\cal A}$ and odd/fermionic part ${\cal H}$ (in which the product vanishes: ${\cal H}\times{\cal H}=0$).  We also explain why this is a natural move if one wants to generalize spectral triples from non-commutative geometry (where the algebras ${\cal A}$ and ${\cal B}$ may be noncommutative) to non-associative geometry (where ${\cal A}$ and ${\cal B}$ may also be non-associative).

In this framework, the standard model of particle physics is described by a spectral triple $\{{\cal A},{\cal H},D\}$ which is, in turn, the product of two spectral triples: a first triple 
$\{{\cal A}_{c},{\cal H}_{c},D_{c}\}$ representing ordinary 4-dimensional spacetime, and a second triple $\{{\cal A}_{F},{\cal H}_{F},D_{F}\}$ describing the finite internal space.  Let us briefly summarize these two triples in turn.

The first triple, describing the 4D spacetime manifold ${\cal M}_{4}$, consists of:
\begin{itemize}
\item the algebra ${\cal A}_{c}=C_{\infty}({\cal M}_{4})$ of smooth functions on ${\cal M}_{4}$;
\item the Hilbert space ${\cal H}_{c}$ of square-integrable Dirac spinors on ${\cal M}_{4}$, on which the elements of ${\cal A}_{c}$ act by pointwise multiplication: $f(x)\psi(x)=\tilde{\psi}(x)$ (for $f\in{\cal A}_{c}$ and $\psi,\tilde{\psi}\in{\cal H}_{c}$); and
\item the ordinary curved-space Dirac operator $D_{c}$ acting on ${\cal H}_{c}$.
\end{itemize}

The second spectral triple, describing the finite internal space, consists of:
\begin{itemize}
\item the algebra ${\cal A}_{F}=M_{3}(\mathbb{C})\oplus \mathbb{H}\oplus \mathbb{C}$, whose elements $a\in{\cal A}_{F}$ have the form $a=(m,q,z)$, where $m\in M_{3}(\mathbb{C})$ is a $3\times 3$ complex matrix, $q\in\mathbb{H}$ is a quaternion, and $z\in\mathbb{C}$ is a complex number;
\item the Hilbert space ${\cal H}_{F}=\mathbb{C}^{96}$, where 96 corresponds to the total number of fermionic degres of freedom in the standard model of particle physics (after including particles and anti-particles, left and right chiralities, right-handed neutrinos, 3 colors, and 3 families);
\item and a $96\times96$ Hermitian matrix $D_{F}$ acting on ${\cal H}_{F}$, which is nothing but the mass matrix for the standard model fermions.
\end{itemize}
We still must explain the representation of ${\cal A}_{F}$ as a left-right-bimodule on ${\cal H}_{F}$.  The appropriate representation is well known \cite{Barrett:2006qq, Chamseddine:2006ep}, but recently Barrett found a nicer way to rewrite it 
\cite{Barrett_private}.\footnote{Barrett's new representation is, in turn, closely related to -- and was inspired by -- an earlier representation found by D'Andrea and Dabrowski \cite{DAndreaDabrowski}.  Barrett's contribution was to notice that ${\cal H}_{F}$ could be elegantly represented as an $8\times8$ matrix (rather than as an $8\times4$ matrix as in  \cite{DAndreaDabrowski}): if one does so, then $J$ becomes ordinary hermitian conjugation, and the algebra is {\it also} represented by an $8\times8$ matrix, whose left and right representations on ${\cal H}_{F}$ is simply given by ordinary matrix multiplication.}  
In this new approach, we represent an element $a=(m,q,z)\in{\cal A}_{F}$ by an $8\times8$ block-diagonal matrix
\begin{equation}
  \label{pi(a)}
  \pi(a)=\left(\begin{array}{cc|cc}
  q & & & \\
  & q_{z_{}} & & \\
  \hline
  & & m & \\
  & & & z \end{array}\right)
\end{equation}
where, on the lower right, we have the $3\times3$ block $m$ and $1\times1$ block $z$; while, on the upper left, the $2\times 2$ block $q$ is the standard representation of the quaternion $q$ as a complex $2\times2$ matrix, and the $2\times2$ block $q_{z}$ is the corresponding representation of $z$ as embedded in $\mathbb{H}$
\begin{equation}
  q=\left(\begin{array}{cc}
  \alpha & \beta \\ -\bar{\beta} & \bar{\alpha} \end{array}\right),\qquad\qquad q_{z}=\left(\begin{array}{cc} z & 0 \\ 0 & \bar{z} \end{array}\right).
\end{equation}
Then, for ${\cal H}_{F}$, we write the fermions (and anti-fermions) in a single standard model generation as an $8\times8$ block-{\it off}-diagonal matrix 
\begin{equation}
  \label{psi}
  \psi=\left(
  \arraycolsep=1.4pt\def\arraystretch{1.2}
  \begin{array}{cccc|cccc}
  &&&& \psi_{11}^{} & \psi_{12}^{} & \psi_{13}^{} & \psi_{14}^{} \\
  &&&& \psi_{21}^{} & \psi_{22}^{} & \psi_{23}^{} & \psi_{24}^{} \\
  &&&& \psi_{31}^{} & \psi_{32}^{} & \psi_{33}^{} & \psi_{34}^{} \\
  &&&& \psi_{41}^{} & \psi_{42}^{} & \psi_{43}^{} & \psi_{44}^{} \\
  \hline
  \bar{\psi}_{11}^{} & \bar{\psi}_{21}^{} & \bar{\psi}_{31}^{} & \bar{\psi}_{41}^{} &&&& \\
  \bar{\psi}_{12}^{} & \bar{\psi}_{22}^{} & \bar{\psi}_{32}^{} & \bar{\psi}_{42}^{} &&&& \\
  \bar{\psi}_{13}^{} & \bar{\psi}_{23}^{} & \bar{\psi}_{33}^{} & \bar{\psi}_{43}^{} &&&& \\
  \bar{\psi}_{14}^{} & \bar{\psi}_{24}^{} & \bar{\psi}_{34}^{} & \bar{\psi}_{44}^{} &&&& 
  \end{array}\right)
\end{equation}
where the 16 components in the upper-right block correspond to the 16 Weyl spinors in a single generation of standard model fermions, while the 16 components in the lower-left block are the corresponding anti-fermions.  The left- or right-representation of ${\cal A}_{F}$ on ${\cal H}_{F}$ is simply given, respectively, by the left matrix product $\pi(a)\psi$ or the right matrix product $\psi\pi(a)$, and the map $J$ from particles to anti-particles is simply $J\psi=\psi^{\dagger}$.  To represent all three generations of fermions, we simply take three copies of this $8\times8$ representation.

Finally, the full spectral triple $\{{\cal A},{\cal H},D\}$ is the product of the triple $\{{\cal A}_{c},{\cal H}_{c},D_{c}\}$ describing the continuous 4D manifold ${\cal M}_{4}$ and the triple $\{{\cal A}_{F},{\cal H}_{F},D_{F}\}$ describing the finite internal space:
\begin{equation}
  \label{product_triple}
  {\cal A}={\cal A}_{c}\otimes{\cal A}_{F},\quad
  {\cal H}={\cal H}_{c}\otimes{\cal H}_{F},\quad
  D=D_{c}\otimes 1_{F}+1_{c}\otimes D_{F}
\end{equation}
where, in the last equation for $D$, "$\otimes$" denotes the appropriate graded tensor product \cite{Farnsworth:2016qbp,Bizi:2018psj}.

How do these ingredients correspond to the physical properties of the standard model?  

The symmetries of the physical theory correspond to the automorphisms of ${\cal B}\!=\!{\cal A}\oplus{\cal H}$ \cite{Farnsworth:2014vva, Boyle:2016cjt}, and the spinor fields ({\it i.e.}\ the elements of ${\cal H}$) inherit transformation properties under these symmetries.  In particular, the usual $SU(3)\times SU(2)\times U(1)$ gauge symmetries come from the {\it inner} automorphisms of ${\cal B}_{F}$, and (since ${\cal A}_{F}$ and ${\cal B}_{F}$ are associative $\ast$-algebras) these are generated by commutators $[\pi(a),\underline{\;\;\;}\,]$, where $\pi(a)$ is any traceless anti-Hermitian element of the form (\ref{pi(a)}).\footnote{Actually, the inner automorphisms of ${\cal B}_{F}$ are generated by commutators $[\pi(a),\underline{\;\;\;}]$ for {\it all} anti-Hermitian $\pi(a)$, not just traceless anti-Hermitian $\pi(a)$.  In the usual NCG approach to the standard model, this "extra" commutator -- with the anti-Hermitian trace part of $\pi(a)$ -- yields an extra (unwanted, anomalous) gauge symmetry which must be removed by hand, and the so-called "unimodularity condition" was introduced precisely for this reason.  One of the advantages of the Jordan reformulation below is that this problem is avoided -- no unwanted anomalous gauge symmetries arise, so no new rules are needed to excise them.}  It is a helpful exercise to check that, if we compute such a commutator with our $8\times8$ spinor $\psi$, we find that under $SU(3)\times SU(2)\times U(1)$ the components of $\psi$ transform into one another precisely like a single family of standard model fermions and anti-fermions, with the identifications:
\begin{equation}
  \label{color_psi}
  \psi=\left(
  \arraycolsep=1.4pt\def\arraystretch{1.2}
  \begin{array}{cccc|cccc}
  &&&& \,{\textcolor{red}{u_{L}^{}}}\, & \,{\textcolor{darkgreen}{u_{L}^{}}}\, & \,{\textcolor{blue}{u_{L}^{}}}\, & \,\nu_{L}^{}\, \\
  &&&& {\textcolor{red}{d_{L}^{}}} & {\textcolor{darkgreen}{d_{L}^{}}} & {\textcolor{blue}{d_{L}^{}}} & e_{L}^{} \\
  &&&& {\textcolor{red}{u_{R}^{}}} & {\textcolor{darkgreen}{u_{R}^{}}} & {\textcolor{blue}{u_{R}^{}}} & \nu_{R}^{} \\
  &&&& {\textcolor{red}{d_{R}^{}}} & {\textcolor{darkgreen}{d_{R}^{}}} & {\textcolor{blue}{d_{R}^{}}} & e_{R}^{} \\
  \hline
  \,{\textcolor{red}{\bar{u}_{L}^{}}}\, & \,{\textcolor{red}{\bar{d}_{L}^{}}}\, & \,{\textcolor{red}{\bar{u}_{R}^{}}}\, & \,{\textcolor{red}{\bar{d}_{R}^{}}}\, &&&& \\
  {\textcolor{darkgreen}{\bar{u}_{L}^{}}} & {\textcolor{darkgreen}{\bar{d}_{L}^{}}} & {\textcolor{darkgreen}{\bar{u}_{R}^{}}} & {\textcolor{darkgreen}{\bar{d}_{R}^{}}} &&&& \\
  {\textcolor{blue}{\bar{u}_{L}^{}}} & {\textcolor{blue}{\bar{d}_{L}^{}}} & {\textcolor{blue}{\bar{u}_{R}^{}}} & {\textcolor{blue}{\bar{d}_{R}^{}}} &&&& \\
  \bar{\nu}_{L}^{} & \bar{e}_{L}^{} & \bar{\nu}_{R}^{} & \bar{e}_{R}^{} &&&& 
  \end{array}\right).
\end{equation}
Here red, green and blue are the three quark colors, and indicate how the three quark columns (or anti-quark rows) transform into one another like a triplet (or anti-triplet) under $SU(3)$.  

In addition to these {\it inner} automorphisms, ${\cal B}_{F}$ also has one other anomaly-free {\it outer} automorphism \cite{Farnsworth:2014vva}, corresponding to $U(1)_{B-L}$, so that each generation of 16 standard model fermions breaks up into the usual six irreducible representations of $SU(3)_{C}\times SU(2)_{L}\times U(1)_{Y}\times U(1)_{B-L}$:
\begin{equation}
  \label{SM_table}
  \begin{array}{c|c|c|c|c} 
    & SU(3)_{C} & SU(2)_{L} & U(1)_{Y} & U(1)_{B-L} \\
    \hline
    q_{L}^{i} & 3 & 2 & +1/6 & 1/3 \\
    \hline
    u_{R}^{i} & 3 & 1 & +2/3 & 1/3 \\
    \hline
    d_{R}^{i} & 3 & 1 & -1/3 & 1/3 \\
    \hline
    l_{L}^{i} & 1 & 2 & -1/2 & -1 \\
    \hline
    \nu_{R}^{i} & 1 & 1 & 0 & -1 \\
    \hline
    e_{R}^{i} & 1 & 1 & -1 & -1 \\
  \end{array}
\end{equation}
where $q_{L}=(u_{L},d_{L})$ and $l_{L}=(\nu_{L},e_{L})$ are the usual left-handed quark and lepton doublets.  

Note that the fermions to have their usual chirality assignments, so that the chirality operator $\gamma$ acts on $\psi$ as follows:
\begin{equation}
  \label{gamma_psi}
  \gamma\psi=\left(
  \arraycolsep=1.4pt\def\arraystretch{1.2}
  \begin{array}{cccc|cccc}
  &&&& \,-{\textcolor{red}{u_{L}^{}}}\, & \,-{\textcolor{darkgreen}{u_{L}^{}}}\, & \,-{\textcolor{blue}{u_{L}^{}}}\, & \,-\nu_{L}^{}\, \\
  &&&& -{\textcolor{red}{d_{L}^{}}} & -{\textcolor{darkgreen}{d_{L}^{}}} & -{\textcolor{blue}{d_{L}^{}}} & -e_{L}^{} \\
  &&&& {\textcolor{red}{u_{R}^{}}} & {\textcolor{darkgreen}{u_{R}^{}}} & {\textcolor{blue}{u_{R}^{}}} & \nu_{R}^{} \\
  &&&& {\textcolor{red}{d_{R}^{}}} & {\textcolor{darkgreen}{d_{R}^{}}} & {\textcolor{blue}{d_{R}^{}}} & e_{R}^{} \\
  \hline
  \,{\textcolor{red}{\bar{u}_{L}^{}}}\, & \,{\textcolor{red}{\bar{d}_{L}^{}}}\, & \,-{\textcolor{red}{\bar{u}_{R}^{}}}\, & \,-{\textcolor{red}{\bar{d}_{R}^{}}}\, &&&& \\
  {\textcolor{darkgreen}{\bar{u}_{L}^{}}} & {\textcolor{darkgreen}{\bar{d}_{L}^{}}} & -{\textcolor{darkgreen}{\bar{u}_{R}^{}}} & -{\textcolor{darkgreen}{\bar{d}_{R}^{}}} &&&& \\
  {\textcolor{blue}{\bar{u}_{L}^{}}} & {\textcolor{blue}{\bar{d}_{L}^{}}} & -{\textcolor{blue}{\bar{u}_{R}^{}}} & -{\textcolor{blue}{\bar{d}_{R}^{}}} &&&& \\
  \bar{\nu}_{L}^{} & \bar{e}_{L}^{} & -\bar{\nu}_{R}^{} & -\bar{e}_{R}^{} &&&& 
  \end{array}\right).
\end{equation}

Once we know how the spinors transform ($\psi\to U\psi$) under all the automorphisms of ${\cal B}$, we can demand that the Dirac operator $D$ should also transform covariantly under these symmetries ($D\to UDU^{-1}$).  In a first course on gauge theory, one checks that, in order to preserve covariance, one must "correct" the partial derivative, $\partial\to\partial+A$, by adding a connection $A$ which transforms in a certain way.  It is a closely parallel exercise to show that, in order to preserve covariance under the automorphisms of ${\cal B}$, one must correct the Dirac operator, $D\to D_{A}$, by adding connection terms which transform in the appropriate way.  But in this case, one finds that, since the initial expression (\ref{product_triple}) for $D$ contains two different terms, one needs to add two different types of connection terms, with two different types of transformation laws: in order to make the first term $D_{c}\otimes 1_{F}$ covariant, we must add the usual connection terms corresponding to the usual Levi-Civita connection and the usual $SU(3)\times SU(2)\times U(1)$ vector bosons; while in order to make the second term $1_{c}\otimes D_{f}$ covariant, we must add connection terms that transform just like the scalar Higgs doublet.  In this approach, all of the bosonic fields arise in this way, and in this sense are neatly unified.

Now consider the action for the standard model (including right-handed neutrinos): we can split it into a fermionic part ({\it i.e.}\ all the terms involving spinor fields) and a bosonic part (all the remaining terms, which do not involve spinor fields).  In the spectral triple approach, the fermionic part of the action ({\it i.e.}\ the fermion kinetic terms, mass terms, Yukawa terms, and gauge couplings) are elegantly captured by the single term
\begin{equation}
  S_{fermionic}=\langle J\psi|D_{A}\psi\rangle\qquad\qquad|\psi\rangle\in{\cal H}.
\end{equation}

What about the rest of the action, $S_{bosonic}$?  The traditional proposal \cite{Chamseddine:1996zu} has been to obtain the remaining terms from the so called "spectral action" $Tr[f(D/\Lambda)]$, where $\Lambda$ is some energy scale, and $f(x)$ is some real function.  But we suspect this expression is incorrect, for both experimental and theoretical reasons: from the experimental standpoint, it predicts relationships between standard model couplings (including gravitational couplings) which disagree with observations, and from the theoretical standpoint, these relations are also not preserved by the known RG flow of the standard model (regarded as an effective field theory).  The problem with finding the correct bosonic action is intimately tied to the issue of finding the right generalization of the exterior algebra of differential forms (since this underlies the prototypical bosonic action: the Maxwell action).  This is problematic in NCG, and another motivation for Jordan geometry is that it seems to have better properties in this regard (see Appendix A), offering the hope of an improved formula for $S_{bosonic}$.

\section{The standard model from Jordan geometry}

In this Section, we will instead suggest taking ${\cal A}_{F}$ to be a certain Jordan algebra, with a certain $8\times8$ representation.

Let us begin with a quick and dirty argument for our new Jordan formulation, based on combining two clues from the previous section's NCG formulation. The first clue is that Barrett's new representation, in which the algebra ${\cal A}_{F}$ and the Hilbert space ${\cal H}_{F}$ are {\it both} represented by $8\times8$ matrices (with ${\cal A}_{F}$ block-{\it diagonal}, and ${\cal H}_{F}$ block-{\it off-diagonal}), fits beautifully with the reformulation advocated in \cite{Boyle:2014wba, Farnsworth:2014vva, Boyle:2016cjt} in which ${\cal A}_{F}$ and ${\cal H}_{F}$ are combined into a single super-algebra ${\cal B}_{F}={\cal A}_{F}\oplus{\cal H}_{F}$ (with even/bosonic part ${\cal A}_{F}$ and odd/fermionic part ${\cal H}_{F}$).  The second clue is that, once we write the spinor $\psi$ in $8\times8$ form, we notice that it seems to have a naturally hermitian character (since, physically, particles and their anti-particles are not independent {\it complex} degrees of freedom, but rather independent real degrees of freedom related by complex conjugation).  But if ${\cal A}_{F}$ and ${\cal H}_{F}$ are both $8\times8$ matrices in the same algebra, and if $\psi\in{\cal H}_{F}$ is Hermitian, then $a\in{\cal A}_{F}$ should be Hermitian as well, and should act on $\psi$ by anti-commutation: $\{a,\psi\}\in{\cal H}_{F}$.

In this quick and dirty argument, we cheated slightly, since the finite Hilbert space ${\cal H}_{F}$ does not really satisfy the Hermiticity condition $J_{F}\psi_{F}=\psi_{F}$ on its own in NCG.  Instead, as pointed out by Barrett \cite{Barrett:2006qq, Besnard:2019lak}, it is the full product Hilbert space ${\cal H}={\cal H}_{c}\otimes{\cal H}_{F}$ which must obey the analogous "Hermiticity" condition $J\psi=\psi$.  Nevertheless, just as before, we are then led to infer that ${\cal A}={\cal A}_{c}\otimes{\cal A}_{F}$ (and hence ${\cal A}_{F}$ itself) should be $8\times8$ Hermitian, and should act on $\psi$ by anti-commutation.  

Meanwhile, we are also led to the same conclusion by a different argument.  Barrett's condition $J\psi=\psi$ is needed \cite{Barrett:2006qq, Besnard:2019lak} in order to resolve the "fermion quadrupling problem" which otherwise afflicts the NCG formulation presented in the previous section \cite{Lizzi:1996vr}.  But this leads to another problem since, in the NCG formulation, the condition $J\psi=\psi$ is actually not compatible with ({\it i.e.}\ not preserved by) the representation of ${\cal A}_{c}\otimes{\cal A}_{F}$ on ${\cal H}_{c}\otimes{\cal H}_{F}$.  This problem would be neatly resolved if, instead, ${\cal A}_{F}$ were an algebra of Hermitian matrices acting on $\psi$ by anti-commutation.

Of course, the NCG representation (\ref{pi(a)}) of ${\cal A}_{F}$ described in the previous section is {\it not} Hermitian.  As explained in the Introduction, we cannot hope to find an algebra of Hermitian operators in NCG (since the composition of two Hermitian operators is not Hermitian), and instead we must look for a Jordan algebra.  Can we find an appropriate Jordan algebra ${\cal A}_{F}$ (and representation) to replace $M_{3}(\mathbb{C})\oplus \mathbb{H}\oplus \mathbb{C}$ in Barrett's $8\times8$ construction?

To answer this question, let us first review the classification of finite-dimensional Euclidean Jordan algebras (over the real numbers): these are the algebras identified by Jordan as suitable to describe quantum mechanical observables.  A Jordan algebra is defined in Appendix A, and a {\it Euclidean} Jordan algebra has the additional property that $a_{1}^{2}+a_{2}^{2}+\ldots+a_{n}^{2}=0$ implies $a_{1}=a_{2}=\ldots=a_{n}=0$ (like a vector in Euclidean space, whose components must all vanish if its norm vanishes).  The finite-dimensional Euclidean Jordan algebras were classified by Jordan, Von Neumann and Wigner \cite{Jordan:1933vh}, and fall into four infinite families, plus one exceptional case:
\begin{align}
  \label{JordanList}
\mathbb{S}_n,& &J_n(\mathbb{R}), & &J_n(\mathbb{C}), & &J_n(\mathbb{Q}), & &J_3(\mathbb{O}),
\end{align}
where $n=1,2,\ldots$, and $J_n(\mathbb{\mathbb{K}})$ denotes the Jordan algebra of Hermitian $n\times n$ matrices, with entries drawn from the normed division algebra $\mathbb{K}$, and with product given by the anti-commutator $\{\;,\;\}$.  (The only four normed division algebras $\mathbb{K}$ are the real numbers $\mathbb{R}$, the complex numbers $\mathbb{C}$, the quaternions $\mathbb{H}$, and the octonions $\mathbb{O}$ \cite{Baez:2001dm}). The `spin factor' algebra $\mathbb{S}_n$ may be defined as the vector space $\mathbb{R}\oplus\mathbb{R}^n$, equipped with the product: $(\lambda_{1},v_1)(\lambda_{2},v_2) = (\lambda_1\lambda_2 + \langle v_1|v_2\rangle,\lambda_1 v_2 + \lambda_2 v_1)$, where $\langle\;|\;\rangle$ is the inner product on $\mathbb{R}^n$; or, equivalently, $\mathbb{S}_{n}$ may be thought of as the algebra generated by the $n$ Euclidean Dirac $\gamma$ matrices $\gamma^{\mu}$, with the anti-commutator product.  Also note the following isomorphisms: $J_{1}(\mathbb{K})\cong\mathbb{R}$ and $J_{2}(\mathbb{K})\cong\mathbb{S}_{{\rm dim}(\mathbb{K})+1}$, where $\mathbb{K}=\mathbb{R},\mathbb{C},\mathbb{H},\mathbb{O}$ and ${\rm dim}(\mathbb{K})=1,2,4,8$, respectively.  Corresponding to the Jordan algebras (\ref{JordanList}), we have the automorphism groups $Spin(n)$, $SO(n)$, $SU(n)$, $Sp(n)$ and $F_{4}$, respectively.

Now we can return to our previous question.  We would like a Jordan algebra ${\cal A}_{F}$ with the following properties: (i) it still can be represented by an $8\times8$ block-diagonal matrix (which now acts by anti-commutator on the $8\times8$ block-off-diagonal matrix of spinors $\psi$); (ii) the corresponding algebra ${\cal B}_{F}={\cal A}_{F}\oplus{\cal H}_{F}$ still has $SU(3)\times SU(2)\times U(1)$ as its inner automorphism group; and (iii) the action of ${\cal A}_{F}$ on ${\cal H}_{F}$ still commutes with the chirality operator $\gamma$, defined in Eq.~(\ref{gamma_psi}).  It is now straightforward to check that, up to isomorphism, there is a unique algebra/representation that does the trick.  We must take ${\cal A}_{F}$ to be the Jordan algebra
\begin{equation}
  {\cal A}_{F}=J_{3}(\mathbb{C})\oplus J_{2}(\mathbb{C})\oplus \mathbb{S}_{2} \oplus \mathbb{R}
\end{equation}
where we take an element $\{c_{3},c_{2},s_{2},r\}\in J_{3}(\mathbb{C})\oplus J_{2}(\mathbb{C})\oplus \mathbb{S}_{2} \oplus \mathbb{R}$ to be represented by the following $8\times8$ block diagonal matrix
\begin{equation}
  \pi(a)=\left(\begin{array}{cc|cc}
  c_{2} & && \\
  & s_{2} & & \\
  \hline
  && c_{3} & \\
  &&& r \end{array}\right).
\end{equation}
Since ${\cal A}_{F}$ is a Jordan algebra, the inner automorphisms of ${\cal B}_{F}$ are not generated by commutators with elements $\pi(a)$, but rather by {\it associators} with pairs of elements $\pi(a)$ and $\pi(a')$ -- {\it i.e.}\ by terms of the form $[\pi(a),\underline{\;\;\;}\,,\pi(a')]=\{\{\pi(a),\underline{\;\;\;}\,\},\pi(a')\}-\{\pi(a),\{\underline{\;\;\;}\,,\pi(a')\}\}$, where $a,a'\in{\cal A}_{F}$.  These generate the desired gauge group, $SU(3)\times SU(2)\times U(1)$.  But when we check how the fermions $\psi$ transform under these symmetries, we seem to run into a problem: with the same labeling as in (\ref{color_psi}, \ref{SM_table}), we find that the fermions break into the following six $SU(3)\times SU(2)\times U(1)$ irreps:
\begin{equation}
  \label{Jordan_table1}
  \begin{array}{c|c|c|c} 
    & SU(3) & SU(2) & U(1) \\
    \hline
    q_{L}^{i} & 3 & 2 & 0 \\
    \hline
    u_{R}^{i} & 3 & 1 & +1/2 \\
    \hline
    d_{R}^{i} & 3 & 1 & -1/2 \\
    \hline
    l_{L}^{i} & 1 & 2 & 0 \\
    \hline
    \nu_{R}^{i} & 1 & 1 & +1/2 \\
    \hline
    e_{R}^{i} & 1 & 1 & -1/2 \\
  \end{array}
\end{equation}
which does not agree with the observed charge assignments in the standard model, corresponding to the first three columns in (\ref{SM_table}).  Fortunately (just as in the NCG formulation \cite{Farnsworth:2014vva,  Boyle:2016cjt}), the outer automorphisms come to the rescue: ${\cal B}_{F}$ has exactly one more anomaly-free outer automorphism corresponding to the gauge symmetry $U(1)_{B-L}$.  When we extend our table (\ref{Jordan_table1}) to include {\it all} the automorphisms of ${\cal B}_{F}$, it becomes
\begin{equation}
  \label{Jordan_table2}
  \begin{array}{c|c|c|c|c} 
    & SU(3) & SU(2) & U(1)_{X} & U(1)_{B-L} \\
    \hline
    q_{L}^{i} & 3 & 2 & 0 & 1/3 \\
    \hline
    u_{R}^{i} & 3 & 1 & +1/2 & 1/3 \\
    \hline
    d_{R}^{i} & 3 & 1 & -1/2 & 1/3 \\
    \hline
    l_{L}^{i} & 1 & 2 & 0 & -1 \\
    \hline
    \nu_{R}^{i} & 1 & 1 & +1/2 & -1 \\
    \hline
    e_{R}^{i} & 1 & 1 & -1/2 & -1 
  \end{array}
\end{equation}
But now we recognize that the third column is just a linear combination of hypercharge and baryon-minus-lepton number, $X=Y+(B-L)/2$, so that this table (\ref{Jordan_table2}) is actually {\it equivalent} to our earlier table (\ref{SM_table}), and is just expressed using a less-familiar basis for the $U(1)\times U(1)$ sector.  

Thus, in Jordan geometry (just as in NCG \cite{Farnsworth:2014vva, Boyle:2016cjt}), we find that the geometry designed to most closely match the standard model of particle physics actually predicts a certain (experimentally viable) {\it extension} of the standard model.  This extension contains the usual set of standard model fermions (including three right-handed neutrinos: one in each generation), but it also includes an extra $U(1)_{B-L}$ gauge symmetry, and two extra bosonic fields: the new $U(1)_{B-L}$ gauge boson, and a new complex scalar fields $\varphi$ which does not couple to $SU(3)\times SU(2)\times U(1)_{Y}$, but {\it does} carry charge $+2$ under $U(1)_{B-L}$ (and is responsible for Higgsing this extra gauge symmetry, so that we don't observe it as an extra long-range force).

\section{The Pati-Salam model from Jordan geometry}

In the previous section, we were trying to find the Jordan geometry that corresponded most closely to the standard model of particle physics, but there is a natural extension that is also worth mentioning. 

Following the same reasoning as before, we are again led to look for a representation $\pi(a)$ of $a\in \mathcal{A}_F$ that is $8\times8$ complex Hermitian.  Then the most {\it general} possibility, which (when acting by anti-commutator on $\psi$) commutes with $\gamma$, is:
\begin{equation}
  \left(\begin{array}{cc|c} 
  c_{2}^{L} && \\ 
  & c_{2}^{R} & \\
  \hline 
  &&\\[-3mm]
  && \;\;\;\; c_{4}\;\;\, \\[2mm]
  \end{array}\right)
\end{equation}
which is a representation of the Euclidean Jordan algebra
\begin{equation}
  {\cal A}_{F}=J_{2}^{L}(\mathbb{C})\oplus J_{2}^{R}(\mathbb{C})\oplus J_{4}(\mathbb{C}),
\end{equation}
where the superscripts "$L$" and "$R$" stand for "left" and "right".   Note that the previous section's finite algebra (and representation) are contained within this section's.

One can check that the automorphism group is now $SU(2)_{L}\times SU(2)_{R}\times SU(4)$, and that the components of the spinor $\psi$ transform under these symmetries like
\begin{equation}
  \psi=\left(\begin{array}{cc|c} 
  && \;\;\;\;\psi_{L}\;\;\;\; \\
  && \;\;\;\;\psi_{R}\;\;\;\; \\
  \hline
  && \\[-2mm]
  \bar{\psi}_{L} & \bar{\psi}_{R} & \\[2mm]
  \end{array}\right)\qquad\qquad\qquad
  \label{PatiSalam_table}
  \begin{array}{c|c|c|c} 
    & SU(4) & SU(2)_{L} & SU(2)_{R} \\
    \hline
    \psi_{L}^{i} & 4 & 2 & 1 \\
    \hline
    \psi_{R}^{i} & 4 & 1 & 2
  \end{array}
\end{equation}
This is a particularly interesting (and experimentally viable) extension of the standard model, the Pati-Salam model \cite{Pati:1974yy, Baez:2009dj}, which is seen to arise very neatly and naturally as a Jordan geometry.  (It has also been argued that the Pati-Salam model appears naturally in the NCG context \cite{Chamseddine:2013rta}.)

\section{Discussion}

Let us end with a few miscellaneous remarks:
\begin{itemize}

\item Over the past several years, there have been a series of intriguing papers \cite{Dubois-Violette:2016kzx, Todorov:2018yvi, Dubois-Violette:2018wgs} which also explore the possibility of using an internal space described by a finite-dimensional Jordan algebra to describe the structure of the standard model.  These papers focus, in particular, on the most interesting Jordan algebra of all: the {\it exceptional} Jordan algebra $J_{3}(\mathbb{O})$.  This is obviously a very appealing possibility, and we are curious to understand how it may relate to our construction.  Another interesting and possibly related construction is in \cite{Furey:2018yyy}.

\item Much work remains on the mathematical side, to flesh out the subject of "Jordan geometry," which -- independent of its physical applications -- seems to offer a rich generalization of classical (commutative-and-associative) geometry.  See \cite{Dubois-Violette:2016kzx, Carotenuto:2018zxd} for some previous work on this topic.  We also note that Lemma 1 in Ref.~\cite{Iochum:1999ih} shows that the distance between pure states in NCG only depends on the Hermitian elements of the algebra, suggesting that notion of distance in NCG should carry over to the Jordan context.  In Appendix A, we describe a natural way to generalize the exterior algebra of differential forms to the Jordan case.  Various other basic issues in Jordan geometry are explored in a follow-up paper \cite{Farnsworth:2020ozj}.

\item Much work remains on the physics side as well, to understand the relationship between Jordan geometry and quantum mechanics, and more generally, to relearn the rules and formalism that describe particle physics and gravity in the Jordan context.

\item As mentioned above, the automorphisms of the coordinate algebra ${\cal A}$ give rise to symmetries of the physical theory.  In particular, if the algebra has the form ${\cal A}=C_{\infty}({\cal M})\otimes {\cal A}_{F}$, where ${\cal A}_{F}$ is a finite-dimensional simple algebra, then the corresponding symmetry group $Aut({\cal A})$ is given by the semi-direct product ${\cal G}\rtimes{\rm Diff}({\cal M})$, where ${\cal G}$ is the group of maps from ${\cal M}$ to ${\rm Aut}({\cal A}_{F})$.  This is precisely the symmetry group of a Yang-Mills theory, with gauge group ${\rm Aut}({\cal A}_{F})$, coupled to gravity.  On the other hand, in attempting to describe the standard model of particle physics or the Pati-Salam model, we are led to use a finite algebra that is the sum of several simple components: {\it e.g.}\ ${\cal A}_{F}=J_{2}^{L}(\mathbb{C})\oplus J_{2}^{R}(\mathbb{C})\oplus J_{4}(\mathbb{C})$ for Pati-Salam.  But in this case, the corresponding symmetry group ${\rm Aut}({\cal A})$ actually contains multiple copies of ${\rm Diff}({\cal M})$.  Could this be a hint that this framework actually predicts multi-metric or multi-tetrad gravity? (See {\it e.g.}\ \cite{Hinterbichler:2012cn} for an introduction to such theories.)

\end{itemize}

\section{Acknowledgements}
We thank Fabien Besnard, Gerard 't Hooft, Michel Dubois-Violette, and especially John Barrett for helpful conversations.  LB acknowledges support from an NSERC Discovery Grant.  Research at the Perimeter Institute is supported by the Government of Canada through the Department of Innovation, Science and Economic Development and by the Province of Ontario through the Ministry of Research and Innovation.  SF thanks the Deutsche Forschungsgemeinschaft for support. 

\appendix

\section{(Graded) Lie and Jordan algebras, and Jordan differential forms}

Lie algebras and Jordan algebras are two sides of the same coin: in the Lie case, the product is the commutator, and in the Jordan case, it is the anti-commutator.  (In the main body of the paper, we write these products as $[a,b]$ and $\{a,b\}$, respectively; but, for notational simplicity, in this Appendix we will write both products simply as $ab$.)  To bring out the parallel, we can take these algebras to be defined by the following three relations:
\begin{subequations}
  \begin{equation}
    \label{ab}
    ab=\left\{\begin{array}{ll}
    +ba&\qquad({\rm Jordan}) \\
    -ba&\qquad({\rm Lie}) \end{array}\right.
  \end{equation}
  \begin{equation}
    \label{abc}
    [a,b,c]+{\rm cyclic}\;{\rm permutations}=0,
  \end{equation}
  \begin{equation}
  \label{Labc}
  (L_{ab}L_{c}-L_{a}L_{bc})+{\rm cyclic}\;{\rm permutations}=0,
  \end{equation}
\end{subequations}
where we are using the standard notation for the associator $[a,b,c]$:
\begin{equation}
  [a,b,c]\equiv(ab)c-a(bc),
\end{equation}
and the left action $L_{a}$
\begin{equation}
  L_{a}x\equiv ax.
\end{equation}
We have written things in terms of these {\it three} relations (\ref{ab}, \ref{abc}, \ref{Labc}) in order to emphasize the relationship between Jordan/Lie algebras, and to make the extension to {\it graded} Jordan/Lie algebras as direct as possible; but it is important to note that in each case, only two of the three relations are actually independent: 
\begin{itemize}
\item in the Jordan case, (\ref{abc}) follows from (\ref{ab}); and 
\item in the Lie case, (\ref{Labc}) follows from (\ref{ab}) and (\ref{abc}).  
\end{itemize}
Also note that, in the Lie case, (\ref{abc}) is the Jacobi identity.

Now we extend the above definition to {\it graded} Lie/Jordan algebras, with the (anti-)commutator replaced by the {\it graded} (anti-)commutator.  The following defining relations are equivalent to the standard ones in \cite{Kac}, but have again been rewritten in a way that emphasizes the Lie/Jordan parallel.  We take the defining relations to be:
\begin{subequations}
  \begin{equation}
    \label{ab_graded}
    ab=\left\{\begin{array}{ll}
      +(-1)^{|a||b|}ba&\qquad({\rm graded\;Jordan}) \\
      -(-1)^{|a||b|}ba&\qquad({\rm graded\;Lie}) \end{array}\right.
  \end{equation}
  \begin{equation}
    \label{abc_graded}
    (-1)^{|a||c|}[a,b,c]+{\rm cyclic}\;{\rm permutations}=0,
  \end{equation}
  \begin{equation}
    \label{Labc_graded}
    (-1)^{|a||c|}(L_{ab}L_{c}-L_{a}L_{bc})+{\rm cyclic}\;{\rm permutations}=0,
  \end{equation}   
\end{subequations} 
where $|a|$ denotes the order (or grading) of $a$.  Again, only two of these three are independent: 
\begin{itemize}
\item in the Jordan case, (\ref{abc_graded}) follows from (\ref{ab_graded}); and 
\item in the Lie case, (\ref{Labc_graded}) follows from (\ref{ab_graded}) and (\ref{abc_graded}).  
\end{itemize}
And, again, in the Lie case (\ref{abc_graded}) is the graded Jacobi identity.

Now let us think about how to define the exterior algebra of differential forms in Jordan geometry.  Let us build up to it by considering three cases in turn:
\begin{enumerate}
\item First consider the standard case where $\mathcal{A}$ (an algebra over the field $\mathbb{F}$) is the ordinary (commutative) algebra of functions on some space.  In this case, there is a standard way to construct the corresponding exterior algebra of differential forms $\Omega \mathcal{A}$: for each element $f\in \mathcal{A}$ we introduce a formal symbol $d[f]$, and then consider the algebra $\Omega \mathcal{A}$ generated by the $f$'s and $d[f]$'s, subject to the following relations
\begin{subequations}
\begin{eqnarray}
  \label{1}
  d[f+g]&=&d[f]+d[g] \\
  \label{2}
  d[\lambda f]&=&\lambda d[f]\;({\rm for}\;\lambda\in\mathbb{F}) \\
  \label{3}
  d[fg]&=&d[f]g+fd[g] \\
  \label{4}
  f d[g]&=&d[g] f \\
  \label{5}
  d[f]d[g]&=&-d[g]d[f]
\end{eqnarray}
\end{subequations}
\item Second consider the case where $\mathcal{A}$ is a non-commutative algebra.  Then we must drop the last two relations (\ref{4}, \ref{5}): in the noncommutative case, they are no longer compatible with the requirement that, when we extend $d$ to an operator on $\Omega \mathcal{A}$, it should square to zero and satisfy the graded Leibniz rule~\cite{Boyle:2016cjt}. For this reason, while NCG generalizations have been developed~\cite{DuboisViolette:1988ir, ConnesBook}, they are significantly different from the original commutative case.

\item Third consider the case where ${\cal A}$ is a Jordan algebra.  Now, since $\mathcal{A}$ is commutative again, we should re-instate relations (\ref{4}, \ref{5}) -- {\it i.e.}\ we can construct $\Omega \mathcal{A}$ using the same five relations (\ref{1} -- \ref{5}) that we used in the original construction.  But now notice that actually, in Case 1, we were also implicitly imposing a sixth relation: associativity.  To generalize the construction of differential forms to the Jordan case, we must simply replace that sixth relation by the associativity relation (\ref{Labc_graded}) that is appropriate for Jordan algebras. Recently a derivation based calculus for Jordan algebras, satisfying all of these properties, was proposed in~\cite{Carotenuto:2018zxd}.   
\end{enumerate}

\end{document}